\documentclass[aps,prd,showpacs,eqsecnum,twocolumn,superscriptaddress]{revtex4-1}
\usepackage{amsmath,amssymb,graphicx,color,ulem}

\begin{document}

\title{High-resolution magnetohydrodynamics simulation of black hole-neutron star merger: Mass ejection 
and short gamma-ray burst}

\author{Kenta Kiuchi}
\affiliation{Yukawa Institute for Theoretical Physics, 
Kyoto University, Kyoto, 606-8502, Japan~} 

\author{Yuichiro Sekiguchi}
\affiliation{Yukawa Institute for Theoretical Physics, 
Kyoto University, Kyoto, 606-8502, Japan~} 
\affiliation{Department of Physics, Toho University, Funabashi, Chiba 274-8510, Japan}

\author{Koutarou Kyutoku} 
\affiliation{Interdisciplinary Theoretical Science (iTHES) Research Group, RIKEN, Wako, Saitama 351-0198, Japan~}

\author{Masaru Shibata}
\affiliation{Yukawa Institute for Theoretical Physics, 
Kyoto University, Kyoto, 606-8502, Japan~} 

\author{Keisuke Taniguchi}
\affiliation{Department of Physics, University of the Ryukyus, Nishihara, Okinawa 903-0213, Japan~}

\author{Tomohide Wada}
\affiliation{Tsukuba University of Technology, 
4-3-15 Amakubo, Tsukuba, 305-8520, Japan~}

\date{\today}

\begin{abstract}
We report results of a high-resolution numerical-relativity simulation for the merger of black hole-magnetized neutron star binaries on Japanese supercomputer ``K''. 
We focus on a binary that is subject to tidal disruption and subsequent formation of a massive accretion torus. 
We find the launch of thermally driven torus wind, subsequent formation of 
a funnel wall above the torus and a magnetosphere with collimated poloidal magnetic field, and high Blandford-Znajek luminosity. 
We show for the first time this picture in a self-consistent simulation. 
The turbulence-like motion induced by the non-axisymmetric magnetorotational 
instability as well as the Kelvin-Helmholtz instability inside the accretion torus 
works as an agent to drive the mass accretion 
and converts the accretion energy to thermal energy, which results in the generation of a 
strong wind. 
By an in-depth resolution study, we reveal that high resolution is essential to draw such a picture. 
We also discuss the implication for the r-process nucleosynthesis, the radioactively-powered transient emission, 
and short gamma-ray bursts.

\end{abstract}

\pacs{04.25.D-, 04.30.-w, 04.40.Dg}

\maketitle


\section{Introduction}
The merger of a black hole (BH) and a neutron star (NS) is one of the most promising sources of gravitational 
waves~\cite{LIGO,VIRGO,KAGRA}. It could be also one of the strongest high-energy phenomena 
in the universe, if the NS is tidally disrupted by the companion BH before the onset of the merger. 
Previous numerical-relativity simulations~\cite{Kyutoku:2011vz,Foucart:2010eq,Foucart:2012vn,Foucart:2011mz,Lovelace:2013vma,Deaton:2013sla,Foucart:2014nda,Foucart:2015vpa,Kyutoku:2010zd,Shibata:2009cn,Kyutoku:2013wxa,Kyutoku:2015gda} have shown that after tidal disruption, a system composed of a BH and a torus is formed. 
This system could be the central engine of short gamma-ray bursts (sGRBs)~\cite{Narayan:1992iy}. 
It is also found that during tidal disruption, an appreciable amount of mass is ejected~\cite{Kyutoku:2013wxa,Kyutoku:2015gda,Deaton:2013sla,Foucart:2014nda}. 
Such ejecta, which should be neutron-rich, will subsequently produce heavy elements via r-process nucleosynthesis~\cite{Lattimer}. The produced unstable heavy elements will subsequently decay, heat-up the ejecta, 
and shine~\cite{Li:1998bw}. Such an electromagnetic signal could be an electromagnetic counterpart 
to detected gravitational waves~\cite{Metzger:2011bv,Piran:2012wd,Rosswog:2012wb,Fernandez:2013tya,Bauswein:2014vfa,Tanaka:2013ixa,Hotokezaka:2013kza,Takami:2014oqa,Kisaka:2014kla}.

These facts motivate us to perform physically reliable numerical-relativity simulation of BH-NS mergers. 
Here, the presence of strong magnetic fields is one of the most characteristic properties of NSs~\cite{Manchester}. 
However, the role of the magnetic fields in their merger process is still poorly known. 

BH-NS mergers can be subject to tidal disruption, for a broad range of the NS compactness, 
the mass ratio of BH to NS, and BH spin~\cite{Kyutoku:2011vz,Foucart:2010eq,Foucart:2012vn,Foucart:2011mz,Lovelace:2013vma,Deaton:2013sla,Foucart:2014nda,Foucart:2015vpa,Kyutoku:2010zd,Shibata:2009cn,Kyutoku:2013wxa,Kyutoku:2015gda}.
An accretion torus, expected to be formed around the remnant BH, is subject to 
the magnetorotational instability (MRI)~\cite{Balbus:1991ay}, and thus, the magnetic field will be amplified. 
Previously, it was difficult to perform high-resolution simulations to resolve the fastest growing mode of 
the MRI because of limited computational resources, although several preliminary 
numerical-relativity simulations have been carried out~\cite{Liu:2008xy,Etienne:2011ea,Etienne:2012te,Chawla:2010sw,Paschalidis:2014qra}.

In this paper, we report the results of our latest general relativistic magnetohydrodynamics (GRMHD) 
simulation for the BH-NS merger performed on Japanese supercomputer K. 
The highest-resolution simulation performed so far together with an in-depth resolution study was done.


\section{Method, initial models, and grid setup.}
Einstein's equation is solved in a puncture-Baumgarte-Shapiro-Shibata-Nakamura 
formalism together with fourth-order finite differencing~\cite{SN,BS,Capaneli,Baker} 
and the GRMHD equations are solved by a high-resolution shock capturing scheme. 
The simulations are performed using a fixed-mesh refinement (FMR) algorithm 
in which each refinement level labeled by $i$ covers the cubic domain of  
$x_{(i)} \in [-N \Delta x_{(i)}, N\Delta x_{(i)}]$ with $\Delta x_{(i)}$ being the grid spacing of 
level $i$. $\Delta x_{(i)}=2\Delta x_{(i+1)} $ and $i=1,2\cdots,$ and $i_{\rm max}-1$ (see Refs.~\cite{Kiuchi:2012,Kiuchi:2014} for details). 
Typically, we set $i_{\rm max}=10$ and the finest grid domain is a $(123~{\rm km})^3$ cube. 
To examine how the result depends on the grid resolution, we change $\Delta x_{\rm imax}=120$m, $160$m, 
$202$m, and $270$m, respectively. We show the grid set-up in our simulations in Table~\ref{tab1}. 
In the highest-resolution run, we use $32,768$ CPUs.

As initial data, we prepare a BH-NS binary in quasi-equilibrium using the method of Ref.~\cite{Kyutoku:2009sp}. 
We model the NS by the Akmal-Pandhalipande-Ravenhall EOS~\cite{APR} 
which is compatible with a maximum neutron star mass $\ge 2M_\odot$
as required by current observational constraints~\cite{Demorest,Antoniadis}. 
We set the NS mass, the mass ratio of 
BH to NS, and the dimensionless aligned spin of BH, $\chi$, to be $1.35M_\odot$, $4$, and $0.75$, respectively. 
The initial orbital angular velocity is $Gm_0\Omega/c^3=0.056$, where $G$ is the gravitational constant, $m_0$ is the sum of BH and NS 
gravitational mass in isolation, and $c$ is the speed of light. 
With these parameters, a massive accretion torus is formed after tidal disruption~\cite{Tanaka:2013ixa,Kyutoku:2015gda}. 

The initial magnetic-field configuration is given in terms of the vector potential as~\cite{Shibata:2005mz}
\begin{align*}
A_j=\left( -(y-y_{\rm NS})\delta^x_j + (x-x_{\rm NS})\delta^y_j\right)A_{\rm b}~{\rm max}(P-P_{\rm c},0)^2, 
\end{align*}
where $x_{\rm NS}$ and $y_{\rm NS}$ 
denote the coordinate center of the NS, $P$ is the pressure, $P_c=P(\rho=0.04\rho_{\rm max})$, and $j=x,y,$ and $z$. 
$\rho_{\rm max}$ is the maximum rest-mass density and 
we set $A_{\rm b}$ 
such that the initial maximum magnetic-field strength is $10^{15}$ G.
Even if we start a simulation with this {\it ad hoc} localized seed magnetic field, the resulting torus surrounding 
the remnant BH becomes in a turbulence-like state and a global magnetic field is naturally formed eventually. 
The magnetic-field strength is chosen so that the wavelength of 
the fastest growing mode of the non-axisymmetric MRI is larger than $10{\Delta x_{\rm imax}}$. 
We note that the resulting turbulence-like state should not depend critically on the initial magnetic-field configuration and 
strength as long as the MRI is resolved~\cite{Balbus:1998ja} 
(see also Ref.~\cite{Obergaulinger:2010gf} for 
the discussion of the dependence of the saturation amplitude on the initial field configuration).
The EOS is parametrized by a piecewise polytrope~\cite{rlof2009} 
and a gamma-law EOS is added during the simulation to capture shock heating 
effects with thermal gamma of 1.8. 
Other choice of thermal gamma could affect the amount of the torus wind.

\section{Results.}
\begin{figure*}[t]
\begin{center}
\includegraphics[width=18cm,angle=0]{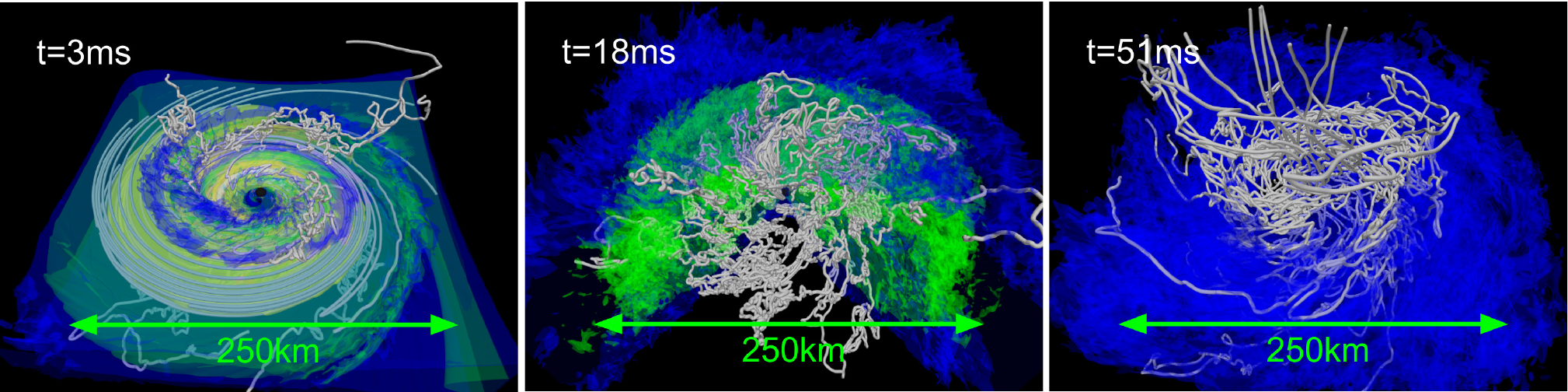}
\end{center}
\caption{\label{fig1}
Snapshots of the rest-mass density profile with the magnetic-field lines 
(a) just after tidal disruption , (b) at an early phase of accretion torus, and (c) in the final phase. 
The iso-surfaces for $10^{11}$, $10^{10}$, and $10^9~{\rm g/cm^3}$ 
are denoted by yellow, green, and blue color. The magnetic-field lines are shown by the white curves. 
In the middle panel, the iso-surfaces are drawn for the three-quarter region. 
}
\end{figure*}

Figure~\ref{fig1} shows snapshots of the rest-mass density profile with the magnetic-field lines. 
Before swallowed by the BH, the NS is tidally disrupted. 
A part of NS matter subsequently forms an accretion torus around the remnant BH 
with mass of $\approx 0.13M_{\odot}$ at $\approx$ 10 
ms after the tidal disruption. 
The Kelvin-Helmholtz instability develops in the contact interfaces 
of the wound spiral arm 
because of the presence of shear motion shown in Fig.~\ref{fig1}a and the vortices are generated subsequently enhancing the magnetic-field energy. 
In addition, the non-axisymmetric MRI activates amplification of the magnetic-field strength~\cite{Balbus:1992} (Fig.~\ref{fig1}b). 
The mass accretion is enhanced by turbulence-like motion which is generated by 
these MHD instabilities as well as by the gravitational torque exerted by the non-axisymmetric 
structure of the accretion torus (see Fig.~\ref{fig1}c and visualization in Ref.~\cite{link}).

\begin{figure}[t]
\begin{center}
\includegraphics[width=8.0cm,angle=0]{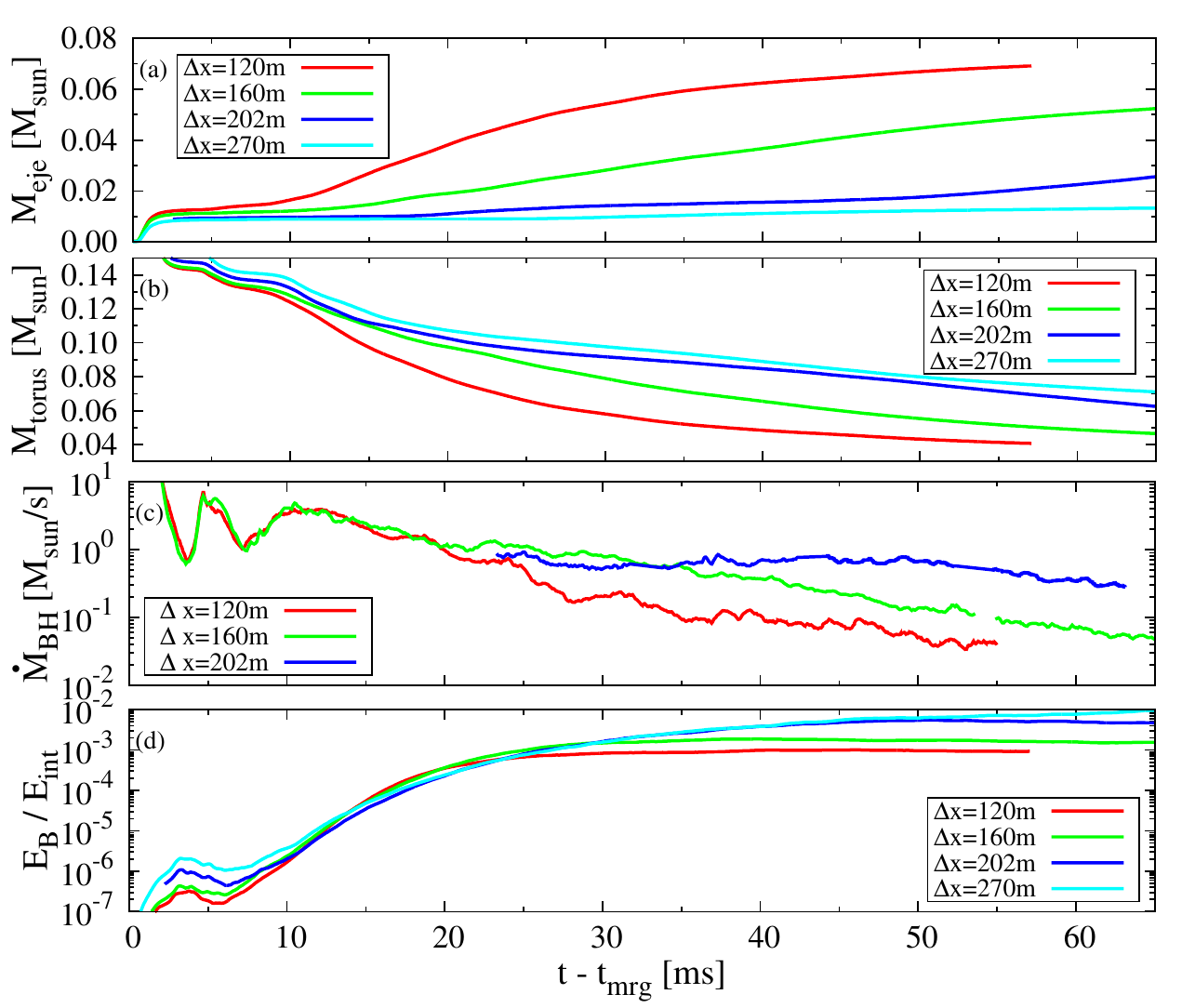}
\end{center}
\caption{\label{fig2} 
Time evolution of (a) ejecta mass, 
(b) torus mass, (c) mass accretion rate onto the BH, 
and (d) magnetic-field energy divided by internal energy.
In panel (c), the data for $t-t_{\rm mrg}\lesssim 20$ ms for $\Delta x=202$ m run is not plotted.
}
\end{figure}

Figure~\ref{fig2} plots the ejecta mass, torus mass, mass accretion rate onto the BH, and 
ratio of the magnetic-field energy, $E_{\rm B}$, to internal energy, $E_{\rm int}$, as functions of time.
We define the merger time $t_{\rm mrg}$ to be the time at which the gravitational-wave amplitude becomes maximal.  
The ejecta is defined to be fluid elements which reside outside the BH and 
have $u_t<-1$ where $u_t$ is the lower time component of the four velocity. 
This criterion means that the fluid elements are gravitationally unbounded. 
The primary mass-ejection mechanism is tidal torque exerted during tidal disruption 
and this {\it dynamical} ejecta is seen for $0~{\rm ms} \lesssim t-t_{\rm mrg} \lesssim 10$ ms in Fig.~\ref{fig2}. 
During this early phase, NS matter of $\sim 0.01 M_\odot$ is ejected approximately along the orbital plane. 
This result agrees with that found in Refs.~\cite{Kyutoku:2013wxa,Kyutoku:2015gda}, 
which demonstrates that our findings will remain robust also for a larger initial separation.

After this primary phase, a new ejecta component appears. 
In the highest-resolution run, the accumulated accretion mass onto the BH for 
$10~{\rm ms}\lesssim t-t_{\rm mrg}\lesssim 30$ ms is $\approx 0.03M_\odot$, while 
the torus mass decreases by $\approx 0.06M_\odot$ over the same time. 
This implies that a significant amount of the torus mass is ejected by a torus wind. 
The launch time and amount of material ejected by the wind depend strongly on the grid resolution: 
The higher-resolution runs result in an earlier launch time and larger amount of the ejecta. 
The mass accretion rate onto the BH also depends on the grid resolution: For higher resolutions, 
it is smaller. The reason for these facts will be described later.

The bottom panel of Fig.~\ref{fig2} shows that irrespective of the grid resolution, 
the magnetic-field energy is exponentially enhanced and eventually saturated: 
$E_{\rm B}$ is typically 0.1\% of $E_{\rm int}$. 
The growth rates of $E_{\rm B}$ for $10~{\rm ms}\lesssim t-t_{\rm mrg}\lesssim 20$ ms 
correspond to 7--8\% of the orbital angular velocity. 
This growth rate agrees approximately with that of the non-axisymmetric MRI predicted by the linear 
perturbation analysis~\cite{Balbus:1992}. 

\begin{figure}[t]
\hspace{-50mm}
\begin{minipage}{0.27\hsize}
\begin{center}
\includegraphics[width=7.5cm,angle=0]{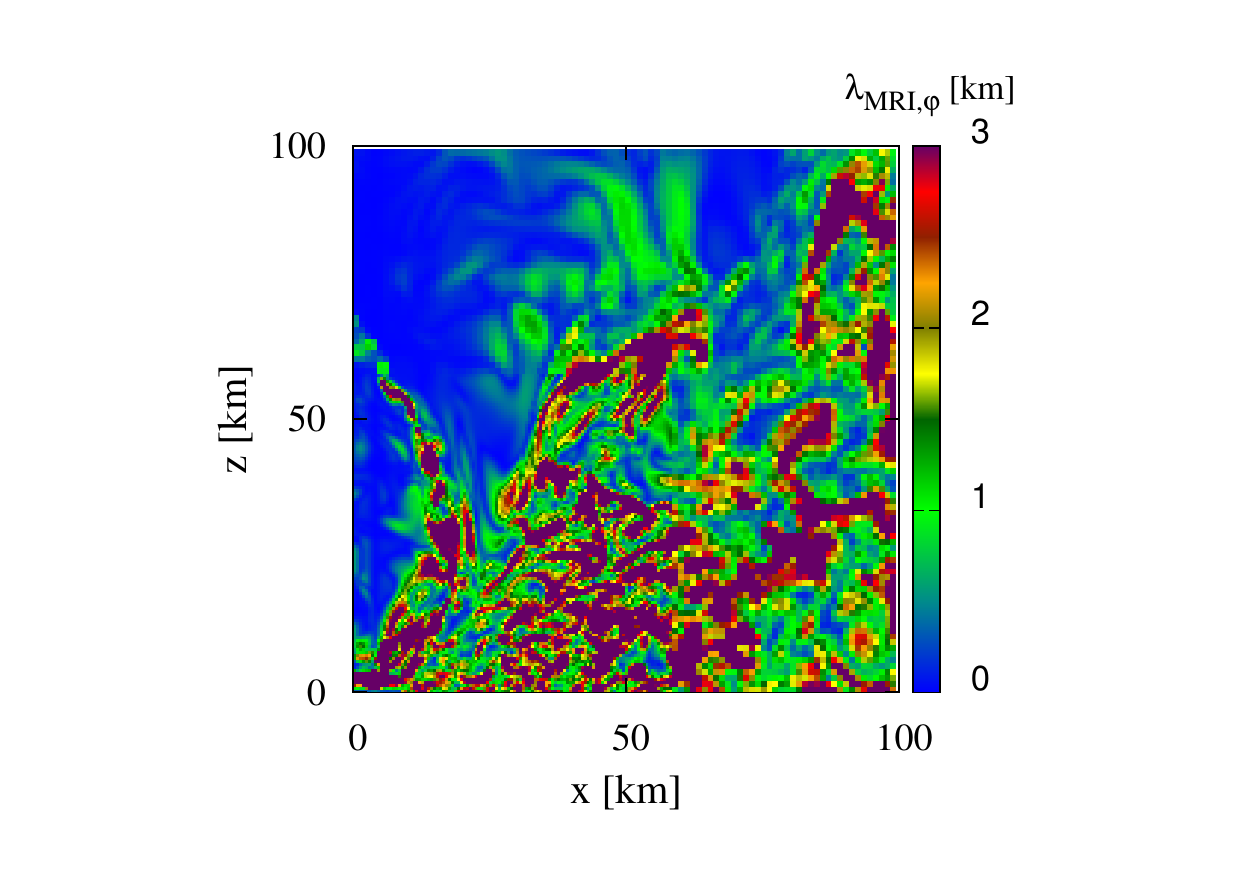}
\end{center}
\end{minipage}
\hspace{15mm}
\begin{minipage}{0.27\hsize}
\begin{center}
\includegraphics[width=7.5cm,angle=0]{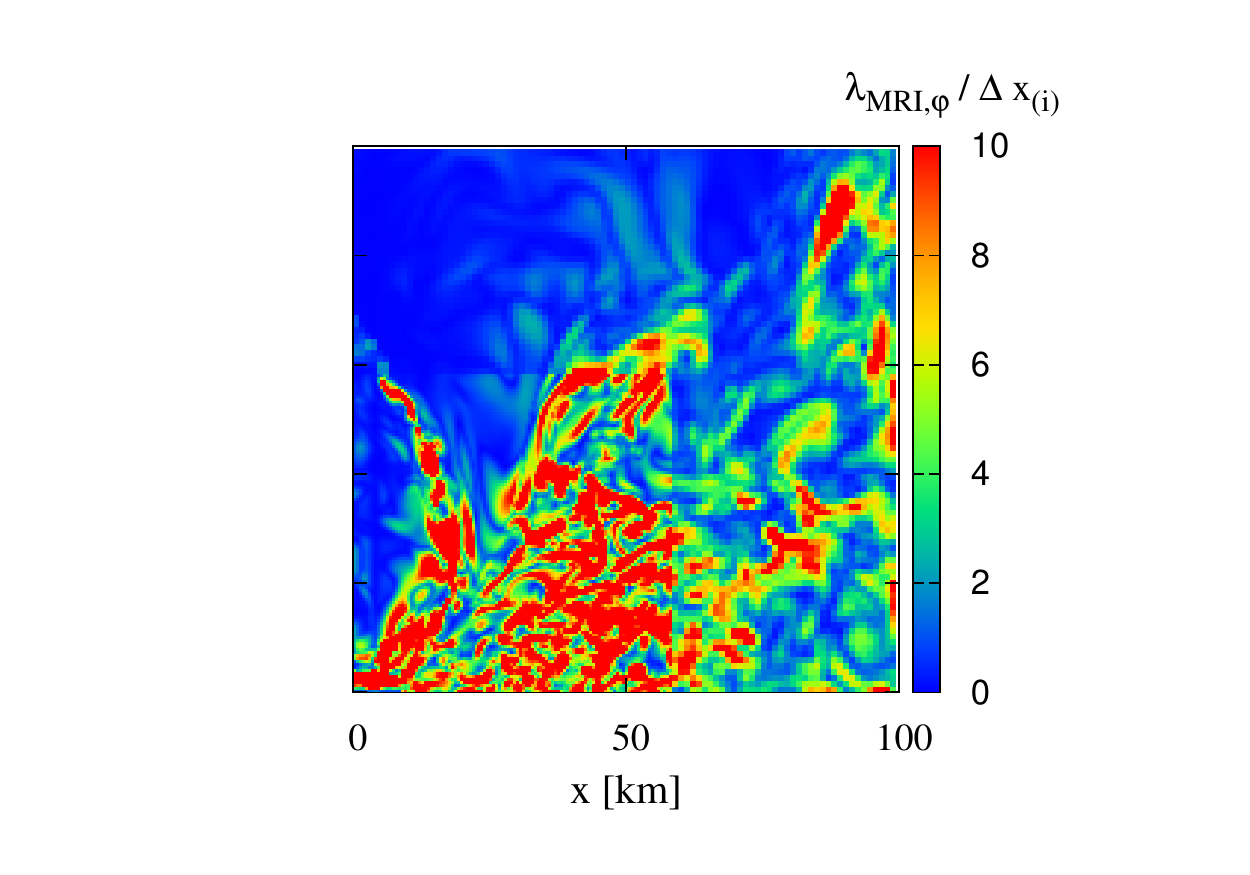}
\end{center}
\end{minipage}
\caption{\label{fig3a}
 Profiles of wavelength of the fastest growing mode for 
 non-axisymmetric MRI (left) 
 and the wavelength divided by the grid resolution (right) 
 on a meridional plane ($x$-$z$) at $t-t_{\rm mrg}\approx 15.0$ ms 
 for the lowest-resolution run.
}
\end{figure}

To see that our grid setting is sufficient to resolve the fastest growing mode of the non-axisymmetric MRI, 
Fig.~\ref{fig3a} plots a snapshot of wavelength of the fastest growing mode of the non-axisymmetric MRI 
on a meridional ($x$-$z$) plane at $t-t_{\rm mrg}\approx 15.0$ ms. We estimate the 
wavelength by 
\begin{align*}
\lambda_{{\rm MRI},\varphi} = \frac{b_{(\varphi)}}{\sqrt{4 \pi \rho h + b^\mu b_\mu}} \frac{2\pi}{\Omega},
\end{align*}
where $b_{(\varphi)}$, $\rho$, $h$, and $\Omega$ are an azimuthal component of 
the magnetic field measured in the fluid rest frame, the rest-mass density, the specific enthalpy, 
and the angular velocity, respectively. 
The wavelength is longer than $\approx 3$ km in a large portion 
of the region and this indicates the fastest growing mode is covered by more than ten grid points even 
in the lowest-resolution run. The right panel of Fig.~\ref{fig3a} clearly shows it. 
Therefore, turbulence-like motion produced by the MRI, which is resolved in our numerical simulation, 
plays an important role in the mass ejection. We discuss this point later. 

In the presence of neutrino radiation, 
the growth rate of the fastest growing MRI mode could be significantly suppressed 
once neutrino viscosity and drag turn on~\cite{Guilet:2014kca}. 
According to Ref.~\cite{Guilet:2014kca}, the neutrino viscosity $\nu$, 
the neutrino mean free path $\lambda_{\nu-{\rm mfp}}$, and the wavelength of the fastest growing MRI mode $\lambda_{{\rm vis-MRI}}$ 
in the viscous regime are : 
\begin{align*}
&\nu = 1.2 \times 10^{12}~T_{10}^2~\rho_{12}^{-2}~{\rm cm^2~s^{-1}},\\
&\lambda_{\nu-{\rm mfp}} = 10^5~\rho_{12}^{-1}~T_{10}^{-2}~{\rm cm},\\
&\lambda_{\rm vis-MRI} = 2.4 \times 10^{5} ~\Omega_3^{-1/2}\nu_{12}^{1/2}~{\rm cm},
\end{align*}
where $T_{10}=T/10{\rm MeV},\rho_{12}=\rho/10^{12}{\rm g cm^{-3}},\Omega_3=\Omega/10^3{\rm s^{-1}},$ 
and $\nu_{12}=\nu/10^{12}{\rm cm^2 s^{-1}}$. 
Note that these estimates would depend on the structure of the accretion torus~\cite{Masada:2006nx,Thompson:2004if}.

Assuming the gas, photons, and 
relativistic electron and positrons contribute to the specific thermal energy, 
we evaluate the temperature from the thermal component of the specific internal energy~\cite{Shibata:2007zm}. 
We utilize the data of the density, angular velocity, and the specific internal energy along $x$-axis 
on the equatorial plane at $t-t_{\rm mrg}\approx 10~{\rm ms}$. 

\begin{figure}[t]
\begin{center}
\includegraphics[width=8.0cm,angle=0]{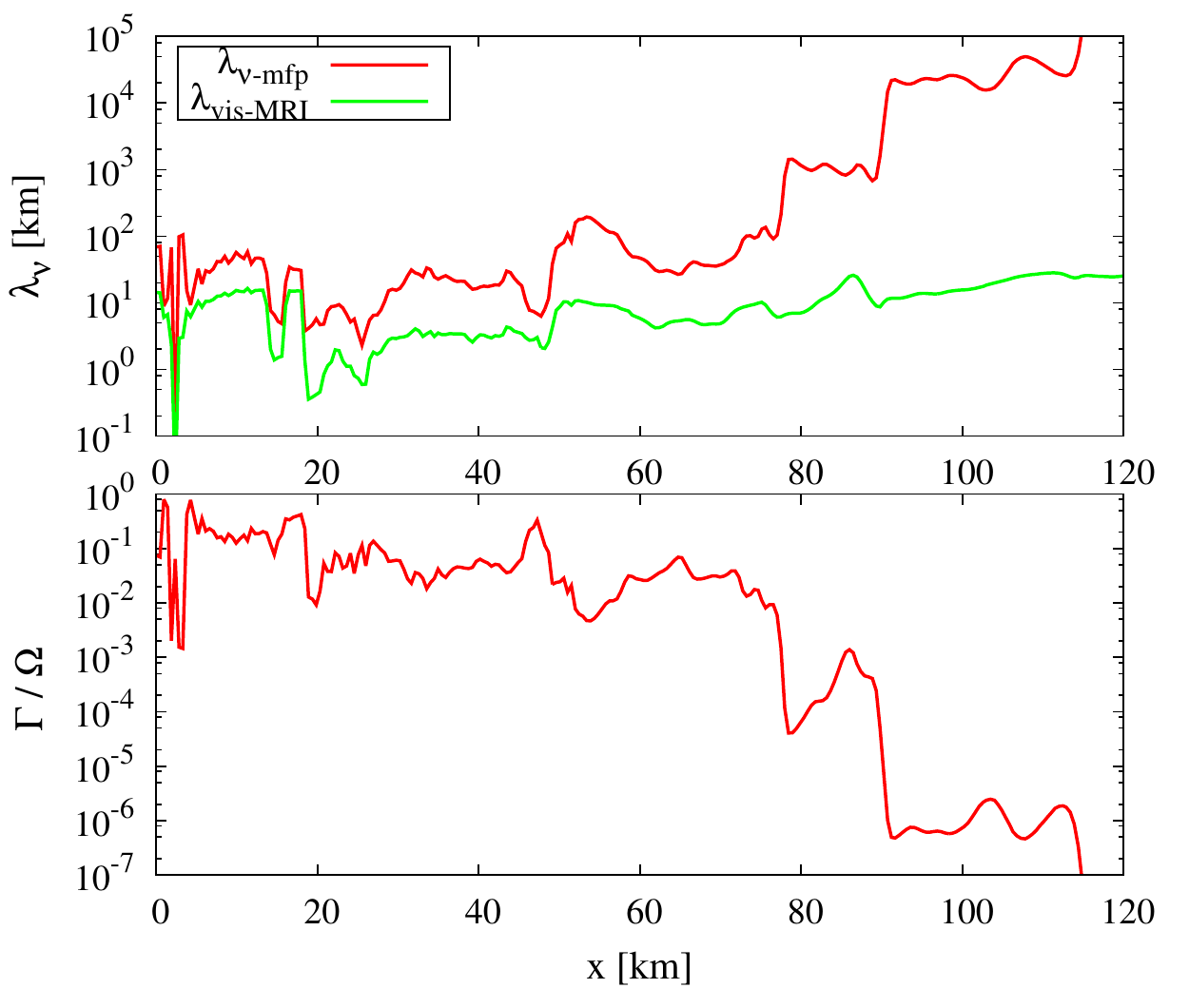}
\end{center}
\caption{\label{fig3b} (Top) radial profile of the neutrino mean free path 
and the wavelength of the fastest growing MRI mode in the viscous region. 
The BH resides at the origin. 
}
\end{figure}

Figure~\ref{fig3b} plots the radial profile of $\lambda_{\nu-{\rm mfp}}$ and $\lambda_{\rm vis-MRI}$. 
In the entire region, $\lambda_{\nu-{\rm mfp}}$ is always longer than $\lambda_{\rm vis-MRI}$, which implies 
that the effect of neutrinos on the MRI is not described by the viscosity, but by the neutrino drag. 
The neutrino drag is characterized by the damping rate $\Gamma$ of the velocity fluctuation due to the 
momentum transport. According to Ref.~\cite{Guilet:2014kca}, $\Gamma=6 \times 10^3T_{10}^6 s^{-1}$. 
The radial profile of $\Gamma$ is shown in Fig.~\ref{fig3b}. Because $\Gamma \ll \Omega$ in 
the entire region, the growth rate of the MRI is not affected by the neutrino drag. 
Although there is an ambiguity in terms of the accretion torus structure and 
the temperature estimation, we conclude that 
the MRI growth rate is not significantly different from that of the ideal MHD in this BH-NS merger. The MRI will grow exponentially even if we assume a weak magnetic-field strength of $\sim 10^{11}$ G.

\begin{figure}[t]
\begin{center}
\includegraphics[width=10.0cm,angle=0]{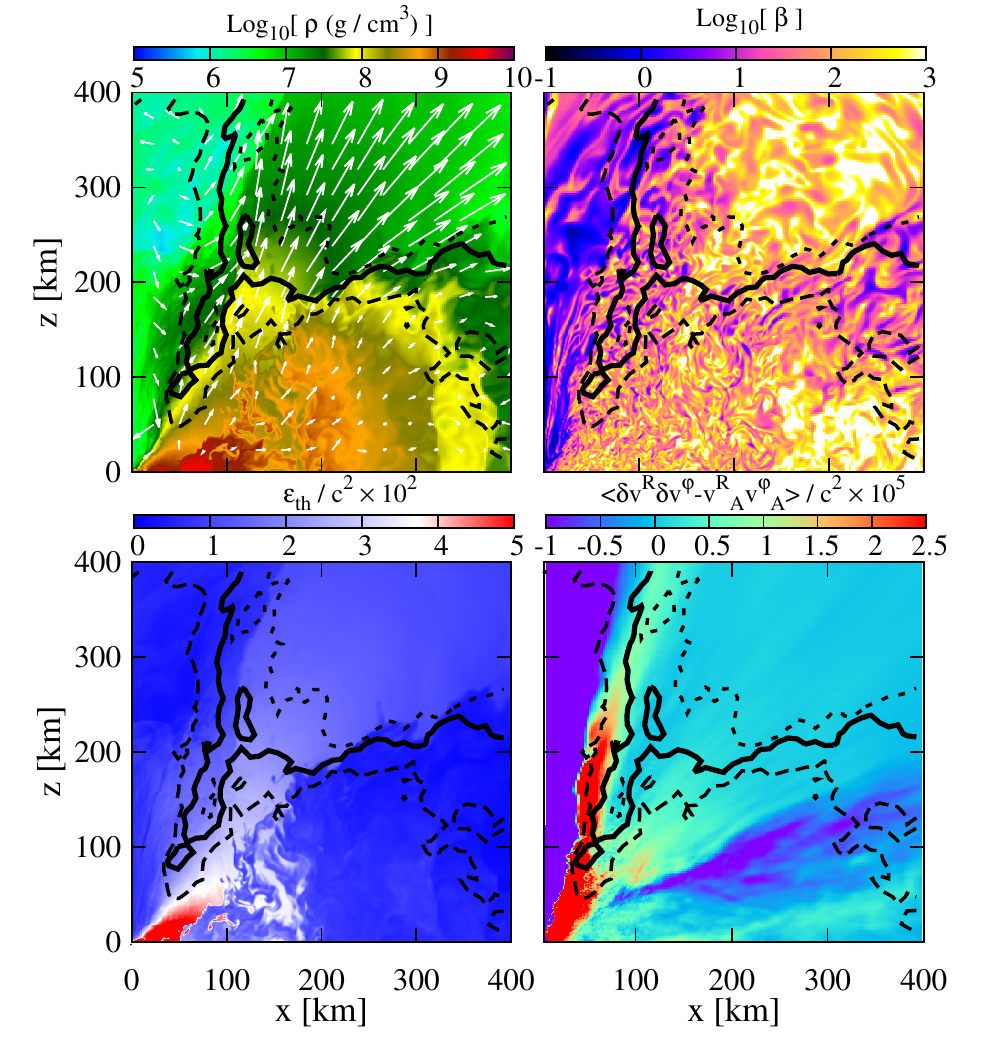}
\end{center}
\caption{\label{fig3}
 Profiles of rest-mass density with velocity fields (top-left), plasma $\beta$ (top-right), 
 thermal component of specific internal energy (bottom-left), 
 and sum of the Maxwell and Reynolds stress (bottom-right) on 
 the meridional plane at $t-t_{\rm mrg}\approx 50.6$ ms for the highest-resolution run. 
 In all the panels, the black curves denote 
 contours with $u_t=-0.98$ (dashed), $-1.00$ (solid), and $-1.02$ (dotted), respectively. 
 In the bottom-right panel, the red (purple) colored region indicates that the energy is 
 transferred outward (inward). 
  } 
\end{figure}

Figure~\ref{fig3} plots snapshots of the rest-mass density, plasma $\beta$ 
(the ratio of matter pressure to magnetic pressure), thermal component of specific internal 
energy, and sum of the Maxwell and Reynolds stress on the $x$-$z$ plane 
at $t-t_{\rm mrg}\approx 50.6$ ms for the highest-resolution run. 
We also plot contours of $u_t$. 
Here, $W \equiv \ln(-u_t)$ is an effective potential of a test particle moving in stationary and axisymmetric BH spacetime~\cite{MTW}. 
The shape of curves in the vicinity of the rotational axis with $u_t=-1$ is approximately 
parabolic. 
In the Newtonian limit, $W$ is reduced 
to $-GM_{\rm BH}/(R^2+z^2)^{1/2}+l^2/2R^2$ where $M_{\rm BH},R$, and $l$ are the BH mass, cylindrical coordinate, 
and specific angular momentum, respectively. 
We assume constant specific angular momentum for simplicity. 
In this case, for a given value of $l$ the contour of $u_t=-1$ becomes parabolic. 
If the specific angular momentum for fluid elements is sufficiently enhanced or fluid elements 
are pushed to high latitude ($z \agt R$) by thermal pressure, they could have $u_t \leq -1$ $(W \ge 0)$~\cite{Hawley:2005xs,Blandford:1998qn,Stone:2000tf} (see the discussion in the next paragraph). 
Because there is no matter in the region above the torus, 
the wind, once it is launched, expands in the widely spread radial direction by contrast with the tidally induced ejecta. 
Subsequently, a funnel wall is formed. 

The point to be clarified is how fluid elements are injected into the region with $u_t<-1$ $(W>0)$. We find that $\beta$ at 
the launch time of the wind is much greater than unity near the torus; 
pure magnetic pressure would not be the main agent of the injection. 
The bottom-left panel of Fig.~\ref{fig3} indicates that there is a 
hot region in the vicinity of the BH, which produces a steep pressure gradient. 
Due to this gradient, the fluid elements are accelerated radially and become unbound once they reach the region 
with $u_t<-1$ $(W>0)$(see Ref.~\cite{Hawley:2005xs} for essentially the same discussion). 

To explore the mechanism to enhance the thermal pressure, we analyze the specific kinetic-energy spectrum $E(k)$, which is calculated by 
$1/2\int\int_V \sum_j e^{-i\vec{k}\cdot\vec{r}}\langle\delta v^j(\vec{x}+\vec{r})\delta v^j (\vec{x})\rangle d^3rd\Omega_k$ where $\vec{k}$ is a wavenumber vector, $k=|\vec{k}|$, $d\Omega_k$ is the volume element in a spherical shell between $k$ and $k+dk$, and $V$ is a cubic region of $x[{\rm km}] \in [50,70]$, $y[{\rm km}] \in [-10,10]$, $z[{\rm km}] \in [-10,10]$. We choose $\vec{x}$ as the center of the cube and $\vec{r}$ is the position vector from the center. 
The fluctuation of the velocity $\delta v^i$ is $v^i-\langle v^i \rangle$ inside the chosen cube. $\langle 
\cdot \rangle$ denotes the time average for the duration $10~{\rm ms} \le t-t_{\rm mrg} \le 20$ ms.

\begin{figure}[t]
\begin{center}
\includegraphics[width=8.0cm,angle=0]{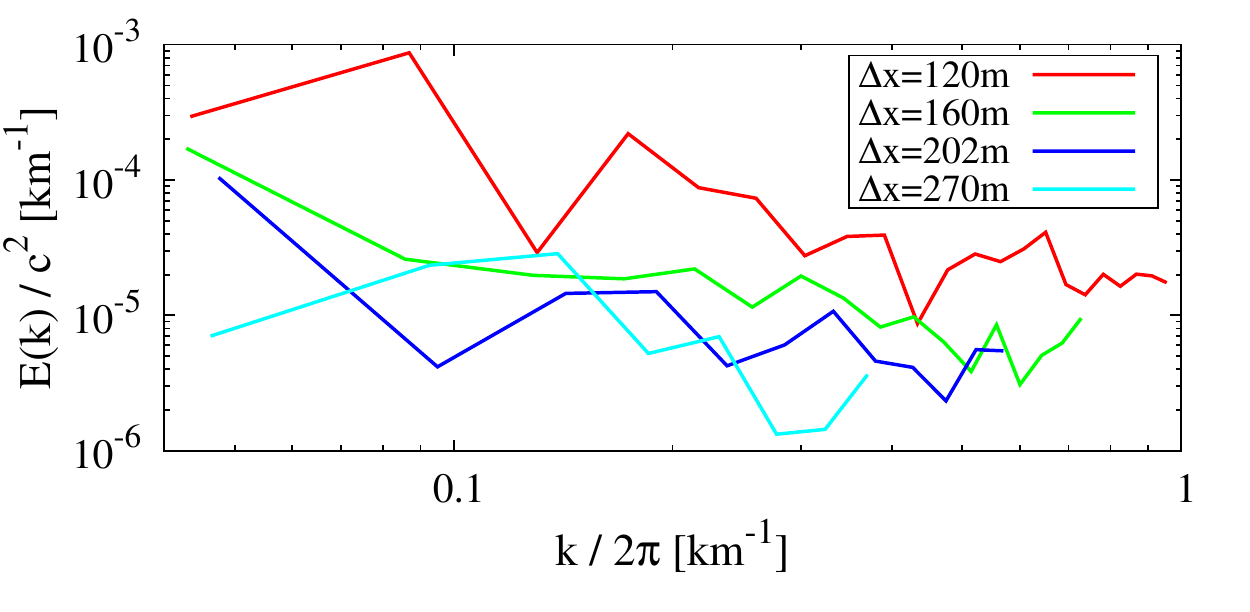}
\end{center}
\caption{\label{fig4}
Energy spectrum of the matter flow in all runs. 
  } 
\end{figure}

Figure~\ref{fig4} plots this energy spectrum for all runs. 
This shows that (i) the spectrum is extended to smaller scales in the 
higher-resolution runs, indicating that the energy is injected at a small scale at which the MRI develops 
and (ii)  the spectrum amplitude is higher for the higher-resolution runs. 
Numerically resolvable wavelength of the fastest growing mode of the MRI becomes shorter with increasing 
the grid resolution. Therefore, the wavenumber for which the MRI develops is given by $k/2\pi=1/10{\Delta x}_{\rm imax}\approx 0.83~{\rm km}^{-1}$ 
for the highest-resolution run, $\approx 0.62~{\rm km}^{-1}$ for the middle-resolution run, $\approx 0.50~{\rm km}^{-1}$ for the normal-resolution run, 
 and $\approx 0.33~{\rm km}^{-1}$ for the lowest-resolution run, respectively. 
This is expected to be approximately equal to the energy injection scale in $E(k)$. 
Assuming that a turbulent state is realized, 
the specific energy dissipation rate is $\sim \delta v^3/l_{\rm edd}$ where $l_{\rm edd}$ is the scale of the vortices~\cite{LL}. 
Because the turbulent energy is proportional to $\delta v^2$, Fig.~\ref{fig4} 
suggests that $\delta v$ becomes higher in the higher-resolution runs for a given scale $l_{\rm edd}$. 
This indicates that the effective turbulent viscosity, $l_{\rm edd}\delta v$, increases with increasing resolution. 
Due to the realistic high viscosity achieved by this turbulence-like motion, 
the mass accretion inside the accretion torus is enhanced in the higher-resolution runs and 
the mass accretion energy is efficiently converted to thermal energy in the vicinity of the BH~\cite{Kuncic:2004qj,Maruta:2003ee,Igumenshchev:2000ki}. 

\begin{figure}[t]
\hspace{-50mm}
\begin{minipage}{0.27\hsize}
\begin{center}
\includegraphics[width=7.5cm,angle=0]{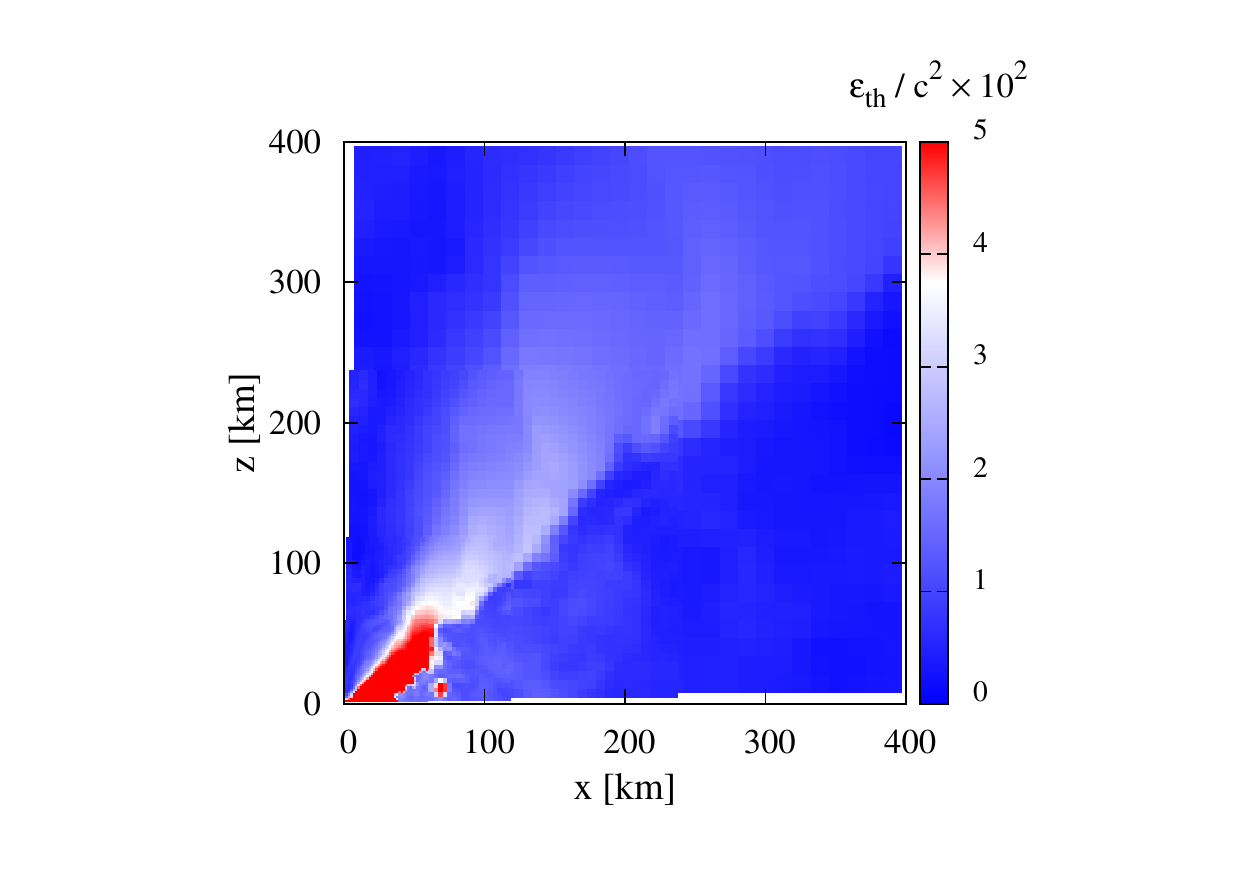}
\end{center}
\end{minipage}
\hspace{15mm}
\begin{minipage}{0.27\hsize}
\begin{center}
\includegraphics[width=7.5cm,angle=0]{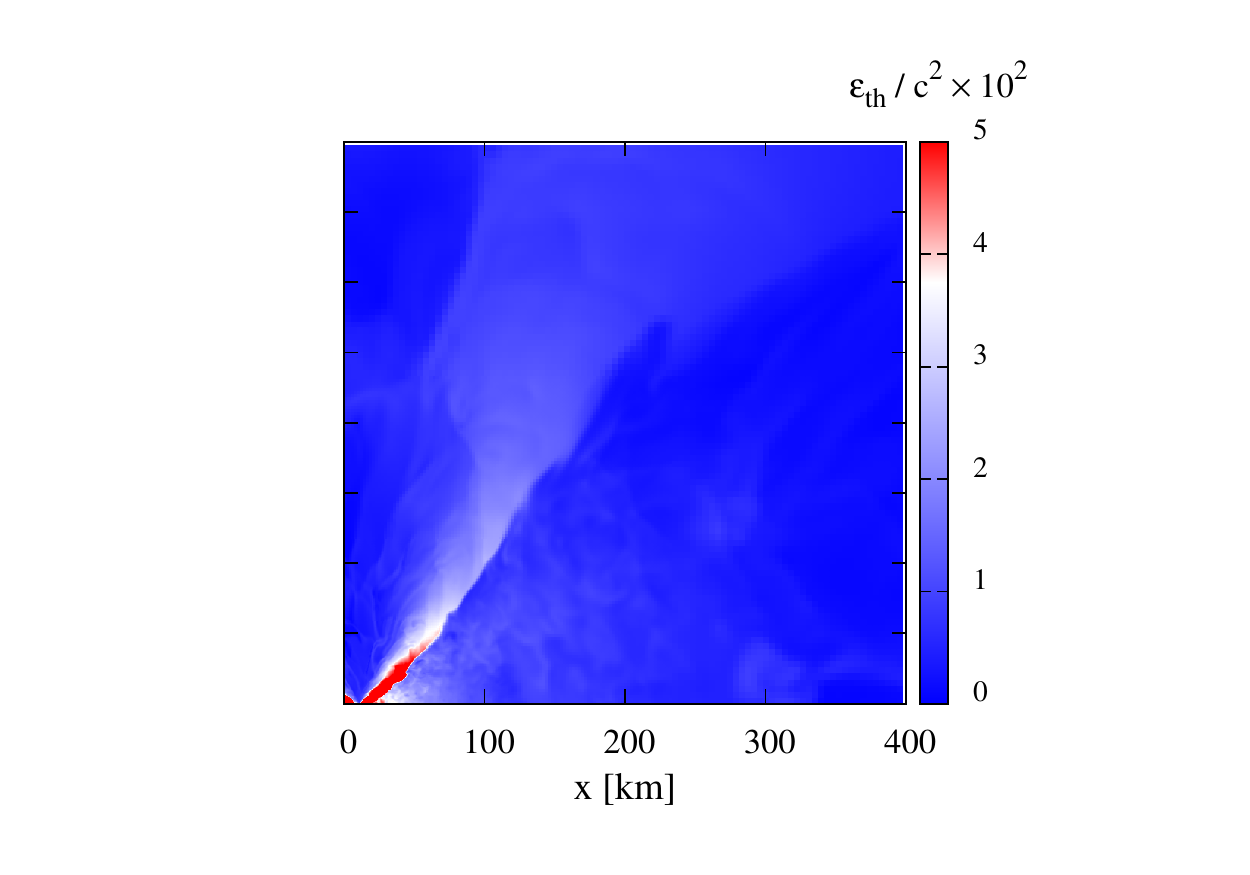}
\end{center}
\end{minipage}
\caption{\label{fig4a}
Profiles of thermal component of specific internal energy for $\Delta x =160$m (left) and $\Delta x=270$m (right) 
on a meridional ($x$-$z$) plane at $t-t_{\rm mrg}\approx 50.6$ ms.
}
\end{figure}

Figure~\ref{fig4a} plots snapshots of the thermal component of specific internal energy on the $x$-$z$ plane 
at $t-t_{\rm mrg} \approx 50.6$ ms for $\Delta x =160$ m and $270$ m. Comparing to the bottom-left panel of Fig.~\ref{fig3}, 
the thermal component of specific internal energy in the vicinity of the inner edge of torus and 
the torus interior is larger than that in the rest of torus in the higher resolution run. 
This trend indicates that efficient thermalization of the mass accretion energy 
due to the effective turbulent viscosity would be realized as discussed above. 
Consequently, the torus-wind power is enhanced while the mass accretion onto the BH is suppressed
~\cite{Kuncic:2004qj,Maruta:2003ee,Igumenshchev:2000ki}, which can be seen clearly in Fig.~\ref{fig2}. 
The bottom-right panel of Fig.~\ref{fig3} indeed shows that the energy is transferred outward. 
In Figs.~\ref{fig2} and \ref{fig4}, the amount of the ejecta mass and the spectrum of the matter flow 
do not exhibit the convergence. However, with the improvement of the grid resolution, the total amount of the torus wind mass 
increases and this indicates that our highest-resolution results would show the lower bound of 
the total wind mass.

The high BH spin with a small horizon radius, which is necessary for tidal disruption of realistic BH-NS binaries~\cite{Kyutoku:2015gda,Foucart:2014nda,Deaton:2013sla,Lovelace:2013vma,Foucart:2011mz,Foucart:2012vn}, prevents 
the fluid elements from being accreted on the remnant BH. Then, the fluid elements tend to stay in 
the vicinity of the BH and the pressure gradient is enhanced~\cite{Shibata:2007}. 
This is also the key to push the fluid elements outward~\cite{Hawley:2005xs}. 
In our simulation, the BH is spun up to $\chi \approx 0.9$, which implies 
the radius of the inner most stable circular orbit is so small
$\approx 2.32 G M_{\rm BH}/c^2$~\cite{Shapiro-Teukolsky} 
that an efficient draining by the BH is prohibited. Hence, the accretion onto the BH is suppressed~\cite{Shibata:2007,Gammie:2003qi} 
(see also the panel (c) of Fig.~\ref{fig2}). 
The amount of the torus wind mass is $\approx 0.06 M_\odot$ in the highest-resolution run, which corresponds to about 50 $\%$ of the torus mass at $t-t_{\rm mrg} \approx 10$ ms. 

\begin{figure}[t]
\begin{center}
\includegraphics[width=8.0cm,angle=0]{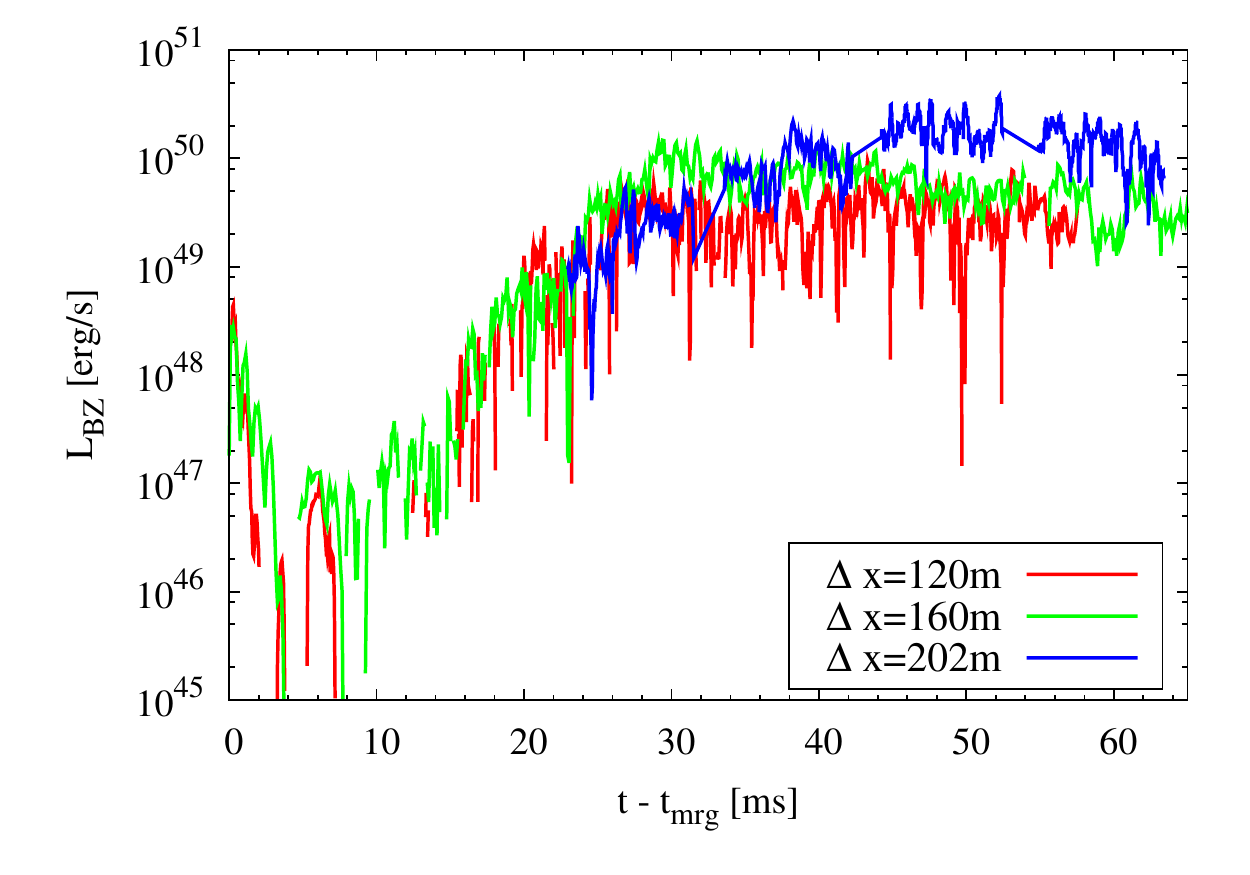}
\end{center}
\caption{\label{fig5}
 Time evolution of outgoing Poynting flux estimated on an apparent horizon. 
}
\end{figure}

As discussed above, the pressure gradient in the vicinity of the BH accelerates the outflow 
and this results in the formation of coherent poloidal magnetic fields 
because the magnetic-field lines are frozen into fluid elements. 
A low-$\beta$ region is formed along the funnel wall in the wind phase. 
Subsequently, the magnetic pressure pushes matter and magnetic-field lines to the polar region 
because there is only dilute matter in this region at the wind launch. 
This results in the formation of a BH magnetosphere. 
The top-right panel of Fig.~\ref{fig3} indeed shows that 
a region with $\beta\sim10^{-1}$ is formed around $z$-axis. 
In the presence of a BH magnetosphere composed of a coherent poloidal magnetic field, the Blandford-Znajek (BZ) 
mechanism~\cite{Blandford:1977ds} efficiently works for the outgoing Poynting flux generation. 
Figure~\ref{fig5} plots time evolution of the outgoing Poynting flux estimated on an apparent horizon. 
This figure shows that the Poynting flux is significantly enhanced after the wind launch because of 
the coherent poloidal magnetic-field formation.The Poynting flux is as high as $\approx 2\times 10^{49}$ erg/s in the end of 
the highest-resolution run.

\begin{table}
\centering
\caption{\label{tab1} 
Grid set up for four different grid-resolution runs. 
$i_{\rm max}$: The number of the refinement levels. 
$\Delta x_{(i_{\rm max})}$ : The grid spacing of the finest refinement level. 
$N$: The grid number in one positive direction. 
}
\begin{tabular}{ccccccccc}
\hline\hline
Model                       &
$i_{\rm max}$                 &
$\Delta x_{(i_{\rm max})}$ [m]  & 
$N$                           \\
\hline
high   & 10 & 120 & 514 \\
middle & 10 & 160 & 378 \\
normal & 8  & 202 & 306 \\
low    & 8  & 270 & 224 \\
\hline\hline
\end{tabular}
\end{table}

\section{Summary and discussion.}
We performed high-resolution GRMHD simulations of a BH-NS merger on the supercomputer K. 
We self-consistently show a series of the processes composed of tidal disruption of the NS, the accretion torus formation, the magnetic-field amplification 
due to the non-axisymmetric MRI, thermally driven torus wind, subsequent formation of the funnel wall and BH 
magnetosphere, and the high BZ luminosity. 

A resolution study revealed that turbulence-like motion works as the agent to drive the mass accretion and convert kinetic energy to thermal energy 
resulting in the generation of a strong wind. 
To show this phenomenon, sufficiently high-resolution simulations are essential. 
After the launch of the torus wind, a funnel wall and magnetosphere with collimated poloidal magnetic fields 
are naturally formed. 

The torus wind and subsequent funnel plus magnetosphere formation have 
the following implications. First, the formed magnetosphere could help launching a jet by the BZ mechanism. 
The high outgoing Poynting flux found in our simulation could be the main engine for 
sGRBs~\cite{Lee:2007js}. 
Also, the jet could be collimated naturally by the pressure exerted by the funnel wall once the wind is launched. 

The torus wind could contribute significantly to r-process nucleosynthesis of heavy elements in BH-NS mergers.
The dynamical ejecta would be neutron-rich and have a low value of electron fraction $Y_e$. 
By contrast, the torus wind component is expected to 
have a higher value of $Y_e$ due to weak interactions because it has high temperature by shock-heating~\cite{Just:2014fka}. 
A mixture of the dynamical and wind components could be a key to reproduce the solar abundance pattern of the r-process 
heavy elements. Note that it was suggested that viscosity-driven and neutrino-driven winds 
from a torus around the BH could reproduce the solar abundances for mass-number 
greater than 90~\cite{Just:2014fka} (see also Refs.~\cite{Wanajo:2014wha,Sekiguchi:2015dma} for the NS-NS case).

Finally, we comment on the kilonova/macronova (radioactively-powered electromagnetic emission) 
model~\cite{Li:1998bw}. 
The amount of the torus wind component in the highest-resolution run is as high as $\approx 0.06M_\odot$ which is 
much larger than that of the dynamical component $\approx 0.01M_\odot$. 
The torus wind would significantly contribute to kilonova/macronova in BH-NS mergers.  
This point should be investigated systematically in a future work.

\acknowledgments
We thank to Christian D. Ott, Kunihito Ioka, and Takeru Suzuki for giving invaluable comments. 
Numerical computations were performed on K computer at AICS, XC30 at
CfCA of NAOJ, FX10 at Information Technology Center of the University of Tokyo, 
and SR16000 at YITP of Kyoto University. 
This work was supported by Grant-in-Aid for Scientific Research
(24244028, 25103510, 25105508, 24740163, 26400267,
15H00783,15H00836,15K05077), for Scientific Research on Innovative
Area (24103001), by HPCI Strategic Program of Japanese MEXT/JSPS (Project numbers hpci130025, 140211, and 150225), 
and by the RIKEN iTHES Research Project. 



\end{document}